\documentclass[]{aa}

\usepackage{amsmath}
\usepackage{booktabs, makecell, tabularx}
\setcellgapes{2pt}
\newcolumntype{L}{>{\raggedright\arraybackslash\linespread{0.84}\selectfont}X}

\newcommand{\hi}{\mbox{H{\small I}}}

\newcommand{\sm}{\ensuremath{\,{\rm M_{\odot}}}\xspace}
\newcommand{\eg}{{e.g.}\xspace}
\newcommand{\mags}{\ensuremath{\,{\rm mag}}\xspace}
\newcommand{\kpc}{\ensuremath{\,{\rm kpc}}\xspace}
\newcommand{\ie}{{i.e.}\xspace}
\newcommand{\Myr}{\ensuremath{\,{\rm Myr}}\xspace}
\newcommand{\pc}{\ensuremath{\,{\rm pc}}\xspace}
\newcommand{\yr}{\ensuremath{\,{\rm yr}}\xspace}

\defcitealias{Blana2020}{B20}
\defcitealias{Blana2024}{B24}
\defcitealias{Adams2018}{AO18}

\usepackage{graphicx}
\usepackage[svgnames]{xcolor}
\usepackage{txfonts}
\begin{document}

   \title{The CHIMERA Survey: The first CO detection in Leo T, the lowest mass known galaxy still hosting cold molecular gas}

    \titlerunning{The CHIMERA Survey: The first CO detection in Leo~T} 

   %\subtitle{I. Overviewing the $\kappa$-mechanism}

\author{V. Villanueva\fnmsep\thanks{These authors contributed equally to this work. Email: vvillanueva@astro-udec.cl, matias.blana.astronomy@gmail.com}\inst{1,15}    
    \and
    Mat\'ias Bla\~na\inst{\text{$\star$}2}
    \and
    Alberto D. Bolatto\inst{3}
    \and
    M\'onica Rubio\inst{4}
    \and
    Elizabeth Tarantino\inst{5}
    \and
    Rodrigo Herrera-Camus\inst{1,15}
    \and 
    Andreas Burkert\inst{6,7,8}
    \and
    Daniel Vaz\inst{9,16}
    \and 
    Justin I. Read\inst{10}
    \and
    Gaspar Galaz\inst{11}
    \and 
    C\'esar Mu\~noz\inst{12}
    \and
    Diego Calder\'on\inst{13}
    \and
    Manuel Behrendt\inst{6,7,8}
    \and 
    Julio A. Carballo-Bello\inst{14}
    \and 
    Emily Gray\inst{10}
    \and
    Michael Fellhauer\inst{1}     
          }

\institute{Departamento de Astronom\'ia, Universidad de Concepci\'on, Avenida Esteban Iturra s/n, Casilla 160-C, Concepci\'on, Chile
\and 
Vicerrector\'ia de Investigaci\'on y Postgrado, Universidad de La Serena,  La Serena 1700000, Chile
\and 
Department of Astronomy, University of Maryland, College Park, MD 20742, USA
\and 
Departamento de Astronom\'ia, Universidad de Chile, Casilla 36-D, Santiago, Chile
\and 
Space Telescope Science Institute, 3700 San Martin Drive, Baltimore, MD 21218, USA
\and 
Max-Planck-Institut f\"ur extraterrestrische Physik, Gie\ss enbachstra\ss e 1, D-85748 Garching bei M\"unchen, Germany
\and
Universit\"ats-Sternwarte, Fakult\"at f\"ur Physik, Ludwig-Maximilians-Universit\"at München, Scheinerstra\ss e 1, D-81679 M\"unchen, Germany
\and
Excellence Cluster ORIGINS, Boltzmannstr 2, D-85748 Garching bei M\"unchen, Germany
\and 
Instituto de Astrof\'isica e Ciências do Espaço, Universidade do Porto, CAUP, Rua das Estrelas, 4150-762 Porto, Portugal
\and 
Department of Physics, University of Surrey, Guildford GU2 7XH, UK
\and
Instituto de Astrof\'isica, Pontificia Universidad Cat\'olica de Chile, Vicu\~na Mackenna 4860, 7820436 Macul, Santiago, Chile
\and     
Departamento de Astronom\'ia, Facultad de Ciencias, Universidad de La Serena. Av. Raul Bitran 1305 , La Serena, Chile
\and 
Max-Planck-Institut f\"ur Astrophysik, Karl-Schwarzschild-Stra\ss e 1, 85748 Garching, Germany
\and 
Instituto de Alta Investigaci\'{o}n, Universidad de Tarapac\'{a}, Casilla 7D, Arica, Chile
\and 
Millennium Nucleus for Galaxies (MINGAL)
\and
Departamento de F\'isica e Astronom\'ia, Faculdade de Ciências, Universidade do Porto, Rua do Campo Alegre 687, PT4169-007 Porto, Portugal
}

   \date{Received May 22 2025 / Accepted July 1 2025}

  \abstract{We report the first CO detection in Leo T, representing the most extreme observation of carbon monoxide molecules in the lowest stellar mass gas-rich dwarf galaxy ($M_{\star}$$\sim$10$^5$ M$_{\odot}$) known to date. We acquired and present new Atacama Compact Array (ACA) $^{12}$CO($J$=1-0) data within our CHIMERA Survey project for the central region of Leo~T, a metal-poor ([M/H]$\sim$-1.7) dwarf in the Milky Way (MW) outskirts. We identified three compact molecular clouds ($<13$ pc) with estimated upper limit virial masses of $M_{\rm mol}$$\sim$5$\times10^{3}$ M$_{\odot}$ each and a total of 1.4$\pm$0.4$\times$10$^{4}$ M$_{\odot}$, corresponding to $\sim\!3\%$ of the total gas mass. 
We obtained CO-to-H$_2$ conversion factors ($\alpha_{\rm CO}$) as high as $\sim$155$\sm ({\rm K\, km\, s^{-1}\, pc^2})^{-1}$ and mean molecular gas surface densities of $\Sigma_{\rm mol}$$\sim$9 M$_\odot$ pc$^{-2}$ that are consistent with values found in dwarf galaxies with extremely low metal content. All CO clouds are shifted ($\sim$60 pc) from the stellar population centers, and only one cloud appears within the densest \hi region. Two clouds have velocity offsets with the \hi of $\Delta v_{\rm los}\sim\!+13$ km s$^{-1}$ being within twice the velocity dispersion ($\Delta v_{\rm los}/\sigma_{\rm \hi,los}\sim2$) and probably bound. However, the northern cloud is faster ($\Delta v_{\rm los}\sim\!+57$ km s$^{-1}$); our models with low halo masses ($M_{\rm h}\! \lesssim \!10^9$ M$_{\odot}$) result in unbound orbits, suggesting that this material is likely being expelled from the dwarf, providing evidence for molecular gas depletion. These properties reveal a perturbed dynamics intertwined with star formation processes in low-mass dwarf galaxies, supporting a scenario of episodic bursts until they are fully quenched by the MW environment.}
 
   {}


   \maketitle
%
%-------------------------------------------------------------------

\vspace{-0.3cm}

\section{Introduction}
\label{sec:intro}

\vspace{-0.3cm}
Dwarf galaxies are the building blocks of massive galaxies and the most abundant galaxy type, and where feedback processes can have very strong effects because of their shallow gravitational potentials.
They have stellar masses below $M_{\star}\! \lesssim\! 10^{9}\sm$ and dark-matter halos masses within $M_{\rm h}\!\lesssim\! 10^{11}\sm$
\citep[e.g., ][]{Revaz2018,Collins2022}.
The smallest dwarfs, which probably inhabit halos within
$M_{\rm h}\!\la\!10^{9}\sm$ \citep{Read2006,Rey2022,Kim2024,Rey2025},
struggle to hold onto their ISM (Sect.\ref{sec:list}) material after a starburst, driving metal expulsion, and even self-quenching 
\citep{Agertz2020,Gray2025}.
For dwarfs, the environment also plays a key role. 
They can lose material or even accrete gas from the IGM and CGM (Sect.\ref{sec:list}) as they move through these media \citep{Ricotti2008,Rey2022}.
If they fall into massive host galaxies or clusters, they can have their ISM stripped away \citep[\eg][]{Mori2000, Gatto2013, Emerick2016}
or experience tidal mass stripping \citep[\eg][]{Read2006a,Fellhauer2007b,Smith2013,Blana2015}.
Furthermore, the metal-poor ISM environments in dwarf galaxies are extreme laboratories to probe the limits of the molecular cloud scaling relations \citep[\eg][]{Bolatto2008, Wong2011, Rubio2015,Filho2016}, and as proxies for high-redshift studies.
Gas-rich dwarf galaxies have very low metallicities and yet have high star formation efficiencies in very compact molecular clouds, such as in the dwarf galaxy WLM$^{Sect.\ref{sec:list}}$ that has dense CO cloud cores detected with ALMA$^{Sect.\ref{sec:list}}$(\citealt{Rubio2015}).
In addition, CO emission in metal-poor ISMs can change the traceability of the molecular gas, resulting in conversion factors that are much higher than in massive galaxies \citep{Bolatto2013a}.

%\vspace{-0.6cm}

In this letter we present new ACA $^{12}$CO($J$=1-0) line emission data in Leo~T as part of the CHIMERA survey. 
This dwarf has a low luminosity \citep[$M_{\rm V}\!=\!-8.0\mags$; ][]{Irwin2007,DeJong2008} \citep[$-7.60\pm0.14\mags$; ][]{Munoz2018}, a stellar mass of $M_{\star}$$\sim$1.4$\times$10$^5$\sm \citep{Weisz2012,Zoutendijk2021}, and is extremely metal-poor with ${\rm [M/H]}\!\sim\!-1.7$ \citep[][\textit{HST}-WFPC2]{Weisz2012}, or ${\rm [Fe/H]}\!=\!-1.53\pm0.05,\,\,\sigma_{\rm Fe/H}\!=\!0.21$ \citep[][MUSE]{Vaz2023}.
It is currently in the outskirts of the MW \citep[$D_{\odot}\!=\!409^{+29}_{-27}\kpc$;][]{Clementini2012}, although it is unclear whether it is on its first in-fall or second 
(\ie, backsplash galaxy; \citealt{Blana2020}, hereafter \citetalias{Blana2020}; \citealt{McConnachie2021}).
This dwarf is \hi gas rich \citep{Ryan-Weber2008} with the atomic gas dominating the baryonic budget with $M_{\rm \hi+He}$$\sim$5.2$\times$10$^5$\sm \citep[][hereafter \citetalias{Adams2018}]{Adams2018}.
Although no massive young stars have been detected in Leo~T, observational evidence suggests recent star formation episodes \citep[<200\Myr,][]{DeJong2008,Vaz2023}.
However, it is unclear whether this galaxy will undergo another star formation event or if it has already stopped forming new stars and is transitioning into a quenched dwarf spheroidal (dSph) galaxy as it sinks into the MW.
The massive \hi reservoir means that the production of m. gas that fuels star formation is a feasible scenario, although until now no CO detections have been reported that could confirm the presence of H$_2$. Here we present the first detection candidates of CO in Leo~T.

\begin{figure}[t]
\centering
\includegraphics[width=0.95\linewidth]{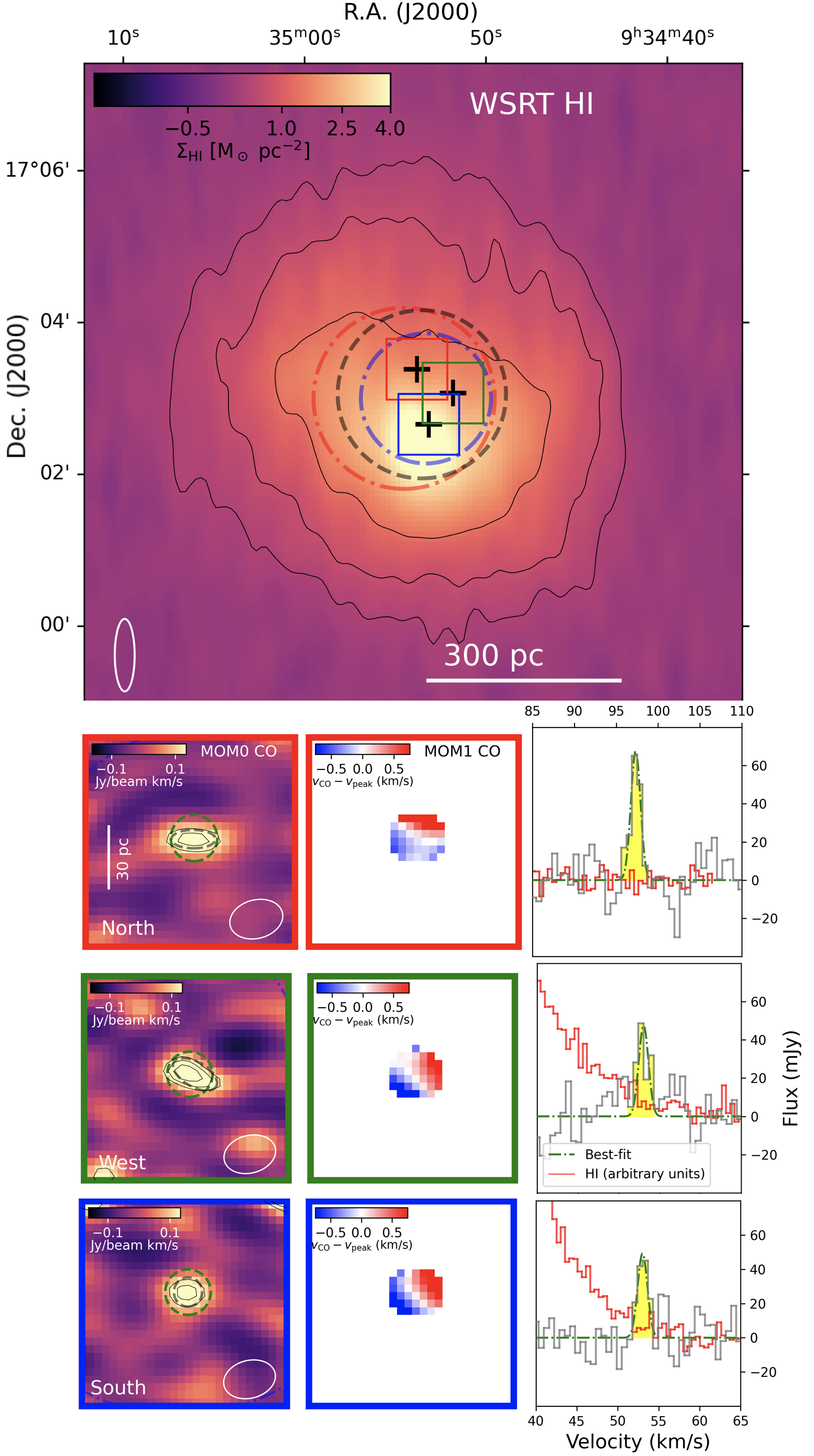}
\vspace{-0.2cm}
\caption{
\textit{Top panel:} Atomic gas surface density map derived from the \hi data \citep{Adams2018}.
The black-dashed circle marks the ACA CO(1-0) data FoV.
Each square marks a cloud detection region: north (red), west (green), and south (blue).
The old and younger stellar population distributions are shown with red and light blue circles (corresponding to their respective half-light radii of 145 pc and 102 pc \citealt{DeJong2008}).
\textit{Bottom nine subpanels:} Zoomed-in images of the ACA CO emission maps (MOM0), velocity fields (MOM1), and spectra of the three CO clouds identified in this work.
The spectra are taken within the apertures shown by the green dashed circles in the MOM0 panels (\ie, $r_{\rm CO}$ in Table \ref{table:obs}). 
The CO and \hi spectra are in heliocentric frame (rightmost subpanels).
The beam sizes are the white ellipses.}
\label{fig:co:map}
\vspace{-0.5cm}
\end{figure}

\vspace{-0.3cm}

\section{Observational data and products} 
\label{sec:obs}
\vspace{-0.3cm}
We used ACA to obtain new CO($J$=1-0) data in Leo~T during the ALMA Cycle 11 program 2024.1.00951.S (P.I. V. Villanueva) within the CHIMERA survey (\textsc{Co H$_2$-Ism Molecular gas ExploRAtion}). This pilot survey aims at conducting a detailed study of the ISM of galaxies at the low-mass ($M_{\star}\!\lesssim \!10^{7}{\rm M}_{\odot}$) and metal-poor ([$Z/{\rm H}$]<-1]) regimes.
CO observations were taken on January 24, 2025, in three execution blocks, spending a total of $\sim$126 minutes on-source. The galaxy was observed with a single pointing using the 7m ACA configuration, with a field of view (FoV) of $\sim$2.2$\times$2.2 arcmin$^2$. We set a spectral bandwidth of $\approx$117 MHz (centered at 115.257 GHz), and a raw spectral resolution of $\sim$0.12 MHz $\approx$0.32 km s$^{-1}$. The scheduling blocks were designed to detect the CO(1-0) emission line down to an rms $\sim$12 mJy/beam at 3 km s$^{-1}$ channel width.

Our ALMA data were calibrated by the observatory pipeline \citep[][version 2024.1.0.8]{Hunter2023}. Both calibration and imaging used {\tt CASA} 6.6.1.17 \citep{CASA2022}. Briefly, the pipeline combined all the $uv$ data for the given target on a common spectral grid, subtracted the continuum emission, and then carried out a deconvolution. The last included a cleaning procedure using the task {\tt tclean}, which was set to stop at 6000 iterations or when a threshold of 10 mJy was reached. The cleaning routine considered a Briggs weighting ({\tt{robust}}=$0.5$; \citealt{Briggs1995}), with a good compromise between the synthesized beam and the signal-to-noise ratio (S/N). This yielded a CO data cube with a beam size $\theta_{\rm min} \times \theta_{\rm maj}=$ $9.''45 \times 13.''18$ (position angle PA=-75$^{\circ}$), a channel width of 0.62 km s$^{-1}$, and a median rms $\sim$20 mJy beam$^{-1}$ (for more details, see Sect.\ref{sec:dataprod}). The resulting angular resolution allowed us to recover physical scales of $\sim$26 pc, with a conversion of $1{\rm [arcsec]}\!=\!1.98$ pc at Leo~T distance.

\vspace{-0.5cm}

\section{Results and discussion}
\label{sec:res}
\vspace{-0.2cm}

Figure\ref{fig:co:map} shows the CO detections of three molecular clouds; the black contours in the intensity maps are the 4.5, 5.5, and 6.5$\sigma$ levels, providing detection reliability. 
To confirm this, we also performed a statistical analysis of the peak S/N per spaxel in the CO field, resulting in the S/N distribution shown in Fig. \ref{fig:peakSNR}.
The spaxels within the detected regions correspond to the red, green and blue bars, which refer to those within the north, west, and south CO clouds; they are well characterized by the highest S/N values of the distribution ($\gtrsim 4.5-6.0$).
The limited angular resolution of the CO(1-0) data does not allow us to fully resolve the sources. When fitting the emission by a 2D Gaussian function, we find full width at half maxima of FWHM$_{\rm CO}$$\sim{10}''$--${14}''$, which are similar to the beam major axis.
We adopted these values to define the physical sizes of the CO structures assuming that they have a spherical shape with a radius $r_{\rm CO} \approx \frac{1}{2}$FWHM$_{\rm CO}$.
We adopted $r_{\rm CO}$ to set a circular aperture and extract the main molecular cloud physical properties (which should be considered as upper limits); they are reported in Table \ref{table:obs}.

We obtained the luminosity of the CO(1-0) line by integrating both its emission within $r_{\rm CO}$ (dashed green circles in bottom left panels of Fig.\ref{fig:co:map}), and in the velocity range defined in Sect.\ref{sec:dataprod} (yellow areas in spectra of Fig.\ref{fig:co:map}). 
Then we used Eq. 1 from \cite{Solomon&VandenBout2005}, after replacing the relevant terms with those of the CO(1-0) emission line; we obtained $L_{\rm CO}({\rm K\, km \,s^{-1}\, pc^{-2}})$=2453$S_{\rm CO}\Delta v D^{2}_{\rm LeoT}$, where $S_{\rm CO}\Delta v$ is the integrated flux density (in Jy km s$^{-1}$), and $D_{\rm LeoT}$ is the distance to Leo~T (in Mpc). We find $L_{\rm CO}$ values in the range of $\sim$32-43 and K km s$^{-1}$ pc$^{2}$ for the CO clouds (Table \ref{table:obs}). Assuming that the CO spectral distribution is primarily driven by molecular gas motions, we computed the virial mass ($M_{\rm vir}$) of each molecular cloud assuming a $1/r$ density law \citep{MacLaren1988}, $M_{\rm vir}({\rm M_\odot})\!=\!\frac{9\sigma^{2} R}{2G}\! \approx\! 1044 R \sigma^{2}$. 
Here, we adopted $R\!\approx\! r_{\rm CO}$ and $\sigma\!\sim\!\sigma_{\rm CO}$ (Table \ref{table:obs}). Upper limit masses of $M_{\rm mol}\!\sim\!5\times10^{3}\sm$ were found for our three CO clouds by adopting $M_{\rm mol}\!\approx\! M_{\rm vir}$ (including heavier elements). 
This adds to a total of $M_{\rm mol,tot}\!=\!1.4\pm0.4\times10^{4}\sm$ and 
corresponds to 2.6\% of the total detected gas (including $M_{\rm \hi+He}$).

Although here we report upper limits,
we also compared the main properties of the three clouds with other molecular cloud studies in the literature, and found consistent agreement.
Detections of CO structures in metal-poor dwarf galaxies (12+log[O/H]<8.0) have been previously attempted (e.g., \citealt{Verter&Hodge1995,Buyle2006}), but only a handful of cases obtained a proper characterization of the CO clouds.
For example, \cite{Rubio2015} reported CO clouds in the extremely metal-poor Local Group dwarf galaxy WLM (12+log[O/H]=7.8; \citealt{Lee2005,Leaman2012}).
They show that CO clouds have luminosities, virial masses, and densities of $\sim$10$^{2}{\rm K\,km\,s^{-1}\,pc^{2}}$, $\sim$2$\times$10$^{3}$ M$_\odot$, and $\sim$6-30 M$_\odot$ pc$^{-2}$, respectively, which are consistent with our results.
However, within our resolution, the Leo~T clouds appear more extended than in WLM, but consistent within the scatter of the main trends observed in other galaxies, as shown in Fig.~\ref{fig:Lco-Mvir}.
We also placed Leo~T in the Kennicutt-Schmidt relation for metal-poor dwarf galaxies, and found a total gas density for each cloud of $\Sigma_{\rm gas}\!\sim\!12\sm\pc^{-2}$, in agreement with other metal-poor dwarf galaxies (e.g., \citealt{Filho2016}; log$[\Sigma_{\rm gas} {\sm\pc^{-2}}] \approx\! 1\!\pm\!1$ in their Fig.1).
Nevertheless, Leo~T's current star formation rate (SFR) remains undetermined, 
where no massive stars, HII regions, or dust have been detected; 
the with H${\alpha}$ upper limit SFR density estimates of $\Sigma_{\rm SFR}<10^{-5}\sm\yr^{-1}\kpc^{-2}$ \citep{Vaz2023}, are much lower than in
other dwarfs (log[$\Sigma_{\rm SFR} {\sm\yr^{-1}\kpc^{-2}}]\, \approx\! -3\pm1$, \citealt{Filho2016}).
However, using historic SFRs derived from stellar population models \citep{Weisz2012} and the observed total cloud areas results in higher values of $\Sigma_{\rm SFR}\!\sim\!10^{-3}\sm\yr^{-1}\kpc^{-2}$, which better fits other dwarfs (see discussion in Sect.\ref{sec:sf}).
Moreover, SFR could be strongly varying if Leo~T has bursty or episodic cycles of active star formation followed by quiescent episodes, as predicted by simulations \citep[\eg][]{Read2016}, where Leo~T would be currently building up a burst of new stars.
Figure\ref{fig:co:map} also shows the central positions of the two main extended stellar populations in Leo~T \citep{Irwin2007,DeJong2008,Vaz2023}: the old population with ages in the range $\sim$5-12 Gyr with a mass of $M_{\star}^{\rm old}$$\sim$$10^5\sm$ and a half-light radius of $r_{\rm h}\!=\!145\pc$ (red circle in Fig.\ref{fig:co:map}), 
and the younger population with ages $<$1 Gyr, $M_{\star}^{\rm y}$$\sim$$10^4\sm$ and $r_{\rm h}=102\pc$ (blue circle).
The global \hi center is close to the old stellar population center; however, the \hi peak is shifted to the south by 80\pc \citepalias[][see also \citetalias{Blana2020} and \citealt{Blana2024} hereafter \citetalias{Blana2024}]{Adams2018}.

Figure\ref{fig:co:map} reveals that the molecular (CO) clouds are spatially shifted from the centers of both stellar distributions and the \hi emission peak by $\sim100\pc$, except for the southern cloud which appears within the dense \hi region.
Such spatial offsets are observed in other dwarf galaxies such as WLM \citep{Rubio2015}, and usually occur where the H{\small I}-to-H$_2$ transition generates a \hi cavity when \hi reaches a critical density of $\Sigma_{\rm crit}$$\gtrsim$10 M$_\odot$ pc$^{-2}$ \citep{Krumholz2009a,Krumholz2013}.
However, here we find values of $\langle\Sigma_{\hi}\rangle$$\sim$2-4 M$_\odot$ pc$^{-2}$ around the CO cloud locations, and even the maximum in Leo~T is 4.9 M$_\odot$ pc$^{-2}$.
Moreover, the observed line-of-sight (los) velocity offsets between the west and south CO clouds with respect to the \hi distribution ($\upsilon_{\rm los,\hi}^{\odot}$=39.6$\pm$0.1 km s$^{-1}$, \cite{Adams2018}) are $\Delta \upsilon_{\rm CO-HI}= \upsilon^{\odot}_{\rm los,CO}\!-\! \upsilon_{\rm los,\hi}^{\odot} \sim$13 km s$^{-1}$ (Fig.\ref{fig:co:map}).
These are within twice the \hi velocity dispersion ($\sigma_{\hi}$=8.3 km s$^{-1}$; \citetalias{Adams2018}), suggesting that they are decoupled from the \hi, but likely gravitationally bound to the dwarf.
Such velocity offsets are not uncommon, as observed in CO clouds in DDO 70 ($\lesssim$20 km s$^{-1}$; \citealt{Shi2020}), 
and are expected for Leo~T's gas supported by random motions, unlike more massive dwarfs where the gas can be rationally supported.

Furthermore, the spatial and velocity gas offsets in Leo~T reveal that its ISM could be perturbed by internal and environmental processes.
HST$^{Sect.\ref{sec:list}}$ observations revealed a population of AGB$^{Sect.\ref{sec:list}}$ star candidates in Leo~T \citep{Weisz2012}, which can have mass outflows of $\dot{M}\!\sim\!10^{-8}-10^{-4}\sm\yr^{-1}$ of slow and cool winds with terminal velocities of $w\sim3-30$ km s$^{-1}$ \citep{Hofner2018},
comparable to the kinematics in Leo~T, which
could be effectively increased by the AGB stars orbital velocities.
Hydro-simulations by \citetalias{Blana2024} show that AGB winds can perturb Leo~T's ISM in low-mass dark-matter halos where the stellar distributions are not fully phase-space mixed.
Moreover, dwarfs in cosmological simulations also show offsets between stars and gas \citep{Rey2022}.
Therefore, a plausible scenario is that stellar winds may compress \hi to reach the critical column density (see Sect.\ref{sec:istab}) and form H$_2$ structures with offset velocities.
Interestingly, we note that the molecular clouds appear to be centered around the younger stellar component (Fig.\ref{fig:co:map}), which could be related to this scenario.

However, the north CO cloud velocity offset is $\Delta \upsilon_{\rm CO-HI}\!=\!+57.7\!\pm0.7$ km s$^{-1}$ (Fig.\ref{fig:North_spectrum}). 
Therefore, we estimated whether the cloud could be gravitationally unbound to Leo~T. 
To test this scenario, we performed orbital calculations with the software \textsc{delorean} \citepalias{Blana2020}; 
with potentials for the stellar, gaseous, and dark matter components, and ram pressure on the cloud from the dwarf's ISM and the MW's CGM with parameters from \citetalias{Blana2024}.
Assuming that the north cloud lies at 60\pc from Leo~T's center and with a speed given only by the los velocity, we can derive a total mass for the dwarf's halo such that the cloud does not leave the dwarf's Jacobi radius due to the MW tidal field.
We find that orbital models with extended cores require total masses greater than $M_{\rm h}>2\!\times\!10^9\sm$ to remain bound.
However, this value can be lower if we consider halos with higher central densities, finding values as low as $M_{\rm h}\!>\!3.5\times10^8\sm$ (model D3 in \citetalias{Blana2024}).
Moreover, while the Leo~T outer layers of less dense \hi can be stripped by the CGM \citepalias[][]{Blana2024}, we find here that ram pressure only weakly affects the clouds' orbits due to the low CGM density, the slow clouds' motions relative to the ISM, and the dense nature of the molecular clouds.
Cosmological simulations with dwarf galaxies show that low-mass dwarf galaxies ($M_{\star}\!\lesssim\!10^6\sm$) inhabit halos with masses of $(1\!-\!5)\!\times\!10^{9}\sm$ \citep[\eg,][]{Rey2022,Rey2025,Gray2025}.
Dynamical models fitted to MUSE-Faint survey observations find a total mass range for Leo~T of $10^{7.88 - 9.23}\sm$ \citep{Zoutendijk2021}, which would allow the north CO cloud to become unbound at the lower end of this mass range.
The cloud's high velocity also suggests that this could belong to foreground MW material. However, this seems unlikely, as we find that the factors $\alpha_{\rm CO}$ would be orders of magnitude higher than typical values for the MW gas (see Sect.\ref{sec:co2h2}).
Therefore, the north cloud kinematics suggest an ongoing expulsion of molecular gas in this low-mass dwarf.

The CO-to-H$_2$ conversion factor ($\alpha_{\rm CO}$) is a useful metric of the physical conditions of the molecular gas since it has been shown to respond to the total surface density of the environment (e.g., \citealt{Bolatto2013a}) and metallicity ($Z$; e.g., \citealt{Amorin2016,Accurso2017}), among others. 
Since $\alpha_{\rm CO}\!=\!M_{\rm mol}/ L_{\rm CO} [{\rm M_\odot}] [{\rm K\, km\, s^{-1}\, pc^2}]^{-1}$ (e.g., \citealt{Leroy2008}), we estimate the cloud $\alpha_{\rm CO}$ (tracing upper limit cloud masses) of CO structures in Leo~T using the virial mass and the luminosity of the CO(1-0) line derived above; we find cloud $\alpha_{\rm CO}$ upper-limit values in the range of 107-156 M$_\odot$ (${\rm K\, km\, s^{-1}\, pc^2}$)$^{-1}$ (see Table \ref{table:obs}). 
These high values of $\alpha_{\rm CO}$, while determined with a limiting resolution, are consistent with estimates of the expected dependence on metallicity, which is extremely low in Leo~T and with a large spread \citep[${\rm [Fe/H]}=-1.53\pm0.05$, $\sigma_{\rm [Fe/H]}\!=\!0.21$; ][]{Vaz2023}. Figure \ref{fig:alpha} includes our derived $\alpha_{\rm CO}$ for CO clouds in Leo~T with other relevant measurements and models from the literature; however, some estimates use various tracers or methods (for more details, see Sect.\ref{sec:co2h2}).
In this context, \cite{Rubio2015,Shi2020} found average core $\alpha_{\rm CO}$ values around of $\sim$30 -- 100 M$_\odot$ (${\rm K\, km\, s^{-1}\, pc^2}$)$^{-1}$ for the CO clouds identified in the dwarfs WLM and DDO 70, where WLM show an extreme cloud value $\alpha_{\rm CO}$$\sim$124 M$_\odot$ (${\rm K\, km\, s^{-1}\, pc^2}$)$^{-1}$ similar to our range.

\vspace{-0.5cm}
\section{Conclusions}
\label{sec:con} 
\vspace{-0.2cm}

We present the first detection of $^{12}$CO($J$=1-0) emission line data in Leo~T, a low-mass ($M_{\star}$$\sim$$10^{5}\sm$) metal-poor dwarf galaxy in the Local Group. Using new CO data taken with the Atacama Compact Array to trace the H$_2$ content, we investigated the main features and properties of the molecular clouds identified in this work. Our main conclusions are the following:
\begin{enumerate}
\item We identified three molecular cloud candidates located in the central region of Leo~T. Within the limited angular resolution of the data beam's semimajor axis of $\sim7$ arcsec ($\sim$13 pc), we computed upper limits for the main physical properties of the molecular clouds (Table \ref{table:obs}).
We find velocity dispersions of $\sim0.62$ km s$^{-1}$, which translates into virial masses of $M_{\rm mol}\sim5\times 10^3\sm$ for each cloud, resulting in a total mass of $M_{\rm mol,tot}\!=\!1.4\!\pm\!0.4\!\times\!10^4\sm$, and mean densities of $\Sigma_{\rm gas}\sim 12\sm\pc^{-2}$.
These results agree with the main properties and scaling relations of molecular (CO) clouds in other low-metallicity dwarf galaxies (see Sect.\ref{sec:res}, \ref{sec:sf} and Fig.\ref{fig:Lco-Mvir}).

\item The CO clouds present spatial and velocity offsets from the stellar and \hi distributions, indicating an unstable dynamical regime. 
Two clouds show low velocities and would remain gravitationally bound to Leo~T, whereas the larger velocity of the north cloud could be direct evidence of ongoing molecular gas ejection in such low-mass dwarfs.
Considering the limited resolution, all CO clouds also show internal rotation.

\item We find upper limits for the CO-to-H$_2$ conversion factor of $\alpha_{\rm CO}$$\sim$107-156 M$_\odot$ (${\rm K\, km\, s^{-1}\, pc^2}$)$^{-1}$. 
These extremely high values are consistent with the very metal-poor nature of Leo~T and with values from environments alike in other dwarf galaxies such as WLM, DDO~70, and DDO~154 (see Sect.\ref{sec:res} and \ref{sec:co2h2}).

\end{enumerate}

Future projects will provide stringent estimates of the properties of the clouds reported in this letter, the current star formation in Leo~T, and cloud properties in other dwarfs, through high-resolution observations, additional CO $J$-transitions, and the characterization of these structures in different environments (as a function of stellar mass, metallicity, and radiation field, among others).

\bibliographystyle{aa}
\bibliography{main}

%%%%%%%%%%

\appendix
\label{sec:app}

\section{ACA-CO(1-0) and WSRT-\hi \, data products}
\label{sec:dataprod}
We used the CO (1-0) data cube to obtain the emission line spectrum (see bottom panels of Fig.\ref{fig:co:map}), centered at the peak of emission and adopting a circular aperture with radius 
$r_{\rm app} = r_{\rm CO} = {\rm FWHM/2}$ (see row 3 in Table \ref{sec:table}).
We also computed the best Gaussian fit to the spectrum in the velocity domain ($v$) to obtain the CO line emission parameters (see the dash-dotted green line in the spectra of the bottom right panels in Fig.\ref{fig:co:map}). Using the model $A e^{-(v-\upsilon)^2 /2\sigma^2}$, we find a mean velocities of $\upsilon^{\odot}_{\rm los, CO}=$97.28$\pm$0.62, 53.03$\pm$1.04, and 53.37$\pm$0.63 km s$^{-1}$ for the north, west, and south CO clouds, respectively.
The best-fit routine also yields velocity dispersions of $\sigma_{\rm CO}=$0.59$\pm$0.62, 0.63$\pm$1.04, and 0.62$\pm$0.63 km s$^{-1}$ for the north, west, and south CO clouds, respectively. We adopted these parameters to derive the CO moment 0 maps ($M_0$; bottom left panels in Fig.\ref{fig:co:map}) by collapsing the data cube in the range $\rm [\upsilon^{\odot}_{\rm los, CO}-FWHM, \upsilon^{\odot}_{\rm los, CO}+FWHM]$, where FWHM is the full width at half maximum (FWHM$=2\sqrt{2\ln(2)}\sigma$) in velocity space. Uncertainties ($u$) of $M_0$s are derived after computing the rms in the signal-free part of the spectra and using $u={\rm rms} \sqrt{N} \Delta v$, where $N$ is the number of channels within the emission and $\Delta v$ is the channel width (in km s$^{-1}$). These calculations also considered the accuracy of the flux calibrator (J0854+2006), which is $\sim$3 mJy. Finally, we compute the moment 1 maps ($M_1$) using Eq. (2) in \citet{Villanueva2024a} and blanking the pixels outside the circular aperture at the coordinates of the emission peaks (see Table \ref{table:obs}) of north, west, and south CO clouds in the $M_0$ (see middle row of 9 bottom panels of Fig.\ref{fig:co:map}).

Similarly as described above, we derive the \hi moment 0 map by collapsing the \hi data cube in the range between $\rm [23.45, 66.21]$ km s$^{-1}$, which corresponds to the spectral range of H{\small I} emission line identified by \citetalias{Adams2018} (see solid red line in bottom left panel of Fig.\ref{fig:co:map}). The final version of the H{\small I} moment 0 map is included in the top-left panel of Fig.\ref{fig:co:map}.
A detailed description of the data can be found in \cite{Adams2018}.

We compare our data with \hi observations of Leo~T taken with the WSRT$^{Sect.\ref{sec:list}}$ telescope by \citetalias{Adams2018}.
The data cube was generated following the standard practice for WSRT data, which consists of calibration and imaging performed on {\tt Miriad} \citep{Sault1995}; the latter was performed by adopting a weighting parameter {\tt robust=0.4}. This procedure produced a data cube with a beam size $\theta_{\rm min} \times \theta_{\rm maj}=$ $15.''7 \times 57.''3$ (or $\sim 31 \pc\times113\pc$ using the conversion of $1{\rm [arcsec]}=1.98\pc$ given the distance to Leo~T of 409\kpc), and a P.A.$= 0.1^{\circ}$. The median rms of the data cube is 0.87 mJy beam$^{-1}$ for the final 0.52 km s$^{-1}$ channel width.

\section{Identification and significance of the CO(1-0) line detections in Leo~T}
\label{sec:SNR}

\begin{figure}[ht!]
\centering
\includegraphics[width=8.5cm]{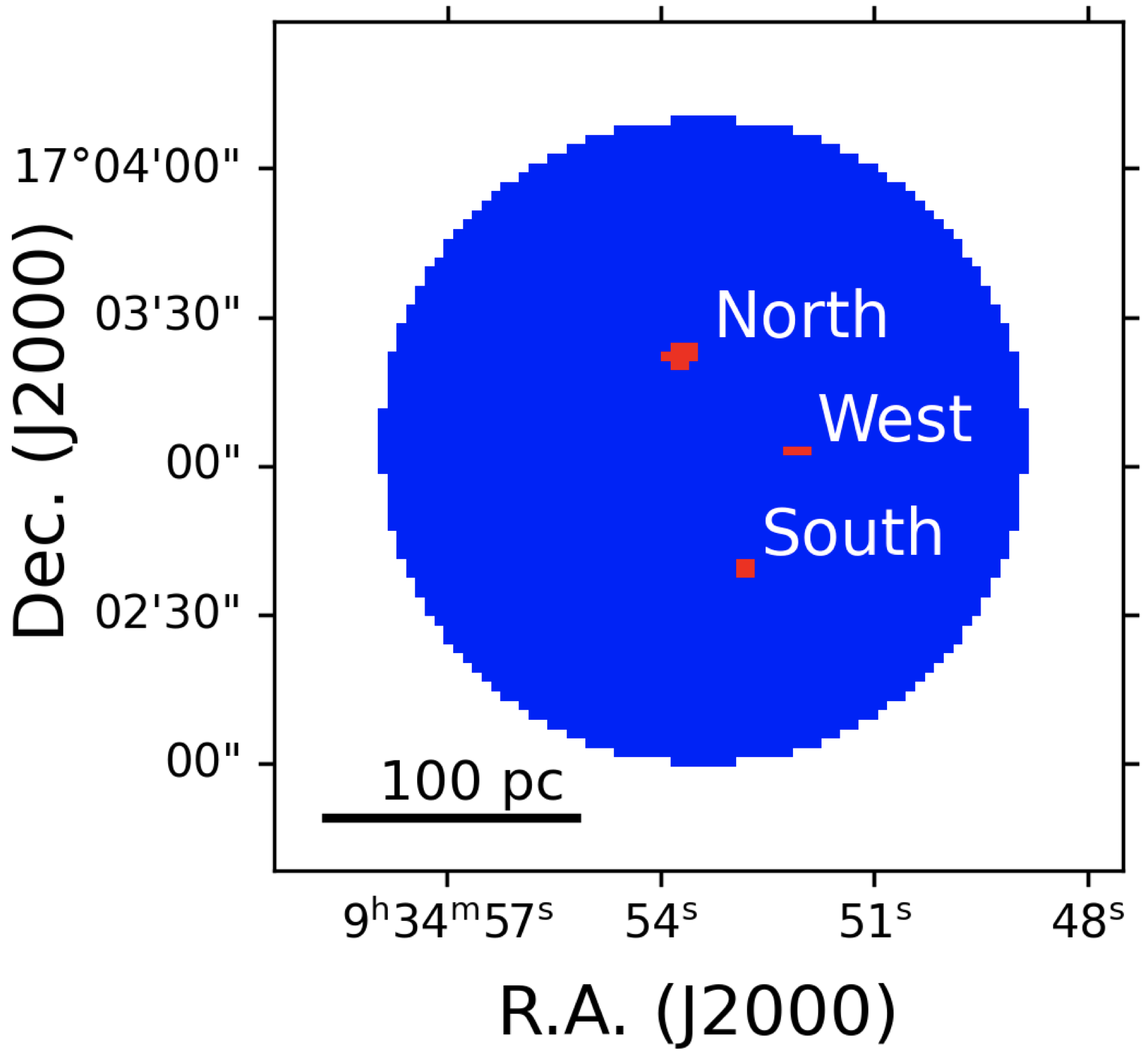}\\
\includegraphics[width=9cm]{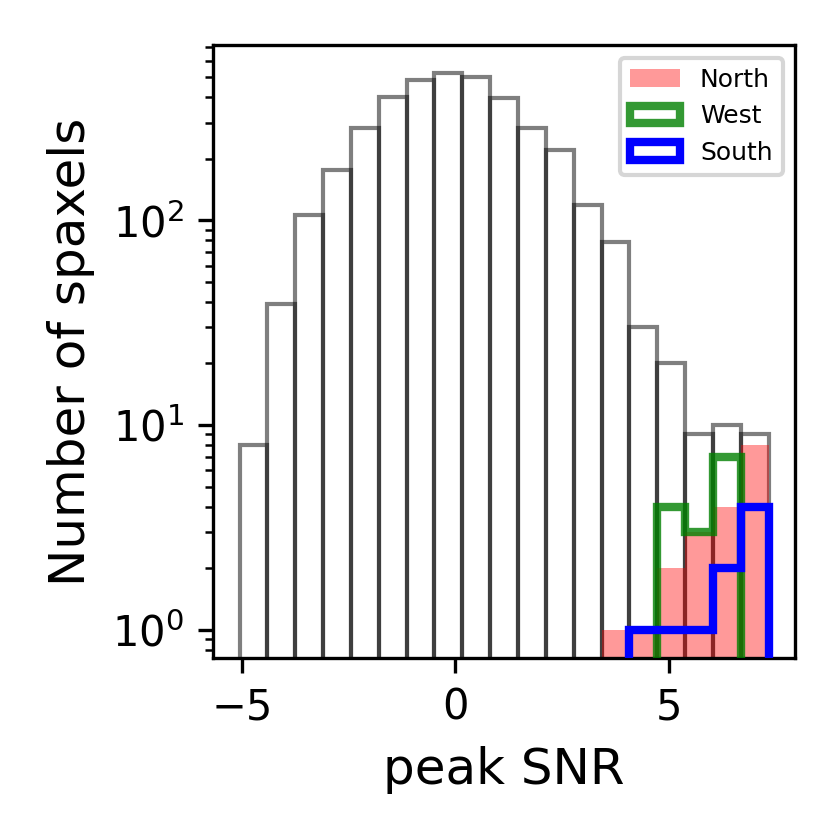}
\vspace{-0.8cm}
\caption{{\it Top:}  Field of view explored with ACA and identification of clouds in the ACA CO(1-0) data cube. Spaxels with one or more channels containing intensities with S/N$\geq5$ are colored red, while those without significant emission (0 spaxels) are colored blue. 
We identify three regions (north, west, south) that include channels with SNR$\geq5$ emissions and also fulfill the requirement of having two or more spaxels meeting the same criteria. 
{\it Bottom:} Histogram of the peak S/N values of the ACA-CO(1-0) data, which resembles a Gaussian distribution (gray). The pixels that are related to the CO detections are identified in the top panel (in addition to all the pixels contained in the aperture of radius $r_{\rm FWHM/2,CO}$) are shown as red, green, and blue bins, corresponding to the north, west, and south CO clouds, respectively.}
\label{fig:peakSNR}
\end{figure}

We performed an identification of cloud-like CO structures in the FoV covered by the ACA-CO(1-0) data presented in this work. To do so, we first compute the root-to-mean square (rms) for all the spaxels within the FoV of the CO(1-0) data. Then, we look for the spaxels that, 1) contain one or more channels with CO emission at least 5 times higher than the rms, and (2) have at least two adjacent spaxels that meet the criteria 1). Selecting spaxels in this way, we identify three regions located in the north, west, and south of the FoV (see top panel of Fig. \ref{fig:peakSNR}), which contain 9, 3, and 4 validated spaxels, respectively (red points).
The bottom panel of Fig. \ref{fig:peakSNR} shows the distribution of positive and negative peak S/N values per pixel in the ACA-CO(1-0) data. The pixels within the region of the CO detections in Leo~T are shown as red, green and blue bins, which are typically well characterized by peak S/N $\gtrsim 4.0-6.0$.

\section{Table with main physical quantities of CO clouds}
\label{sec:table}
We show in Table \ref{table:obs} the main parameters of the CO molecular clouds.
\begin{table}
\caption{Main physical quantities of the CO clouds in Leo~T.}
\vspace{-0.2cm}
\resizebox{1.\linewidth}{!}{\begin{tabular}{lrrr}% <---
\hline
Property Cloud  &  North & West & South  \\
\hline
(1) R.A. & 09$^{\rm h}$:34$^{\rm m}$:53$^{\rm s}$.8 & 09$^{\rm h}$:34$^{\rm m}$:51$^{\rm s}$.8 & 09$^{\rm h}$:34$^{\rm m}$:53$^{\rm s}$.2 \\ 
(2) Dec. & 17$^{\circ}$:03':23''.0 & 17$^{\circ}$:03':04''.2 & 17$^{\circ}$:02':39''.4 \\ 
(3) $r_{\rm CO}(={\rm FWHM/2})$ \,[arcsec] & 5.7$\pm$0.25 & 6.2$\pm$0.8 & 7.1$\pm$0.23 \\ 
(4) $\upsilon^{\odot}_{\rm los,CO} \,[{\rm km\,s^{-1} }]$ & 97.28$\pm$0.62 & 53.0$\pm$1.04 & 53.37$\pm$0.63 \\ 
(5) $\sigma_{\rm CO}\,[{\rm km\,s^{-1}}]$ & 0.59$\pm$0.62 & 0.63$\pm$1.04 & 0.62$\pm$0.63 \\ 
(6) $L_{\rm CO}\,[{\rm K\, km\, s^{-1} pc^{2} }]$ & 43.1$\pm$5.4 & 32.0$\pm$4.4 & 35.5$\pm$4.7 \\ 
(7) $M_{\rm mol}\,[{\rm M_{\rm \odot}}]$ & (4.6$\pm$1.4)$\times10^3$ & (5.0$\pm$1.6)$\times10^3$ & (4.9$\pm$1.5)$\times10^3$ \\  
(8) $\alpha_{\rm CO}\,[{\rm M_{\odot}\, pc^{-2} [K\, km\, s^{-1}]^{-1}}]$ & 107$\pm$36 & 156$\pm$52 & 138$\pm$46 \\ 
(9) $\Sigma_{\rm mol}\,[{\rm M_{\odot}\, pc^{-2}}]$ & 8.6$\pm$0.9 & 9.3$\pm$1.0 & 9.2$\pm$1.0 \\ 
\hline
\end{tabular}}
\vspace{-0.3cm}
\tablefoot{(1) and (2): Coordinates of the pixel with maximum intensity within each CO cloud (J2000). (3) Aperture radius $r_{\rm CO}$ defined as the half of the full width half maximum (FWHM) derived from a 2D Gaussian fit of the emission in MOM1 maps (see Fig.\ref{fig:co:map}) divided by two.
(4) Peak velocity of the CO spectra, (5) velocity dispersion. (6) CO luminosity. (7): Virial mass. (8): CO-to-H$_2$ conversion factor. (9): mean surface mass densities.
(4) to (9) used $r_{\rm FWHM,CO}$, so they should be considered upper limits.}
\label{table:obs}
\end{table}

\section{Scaling and star formation relations in Leo~T}
\label{sec:sf}
\begin{figure}[ht!]
\centering
\includegraphics[width=8cm]{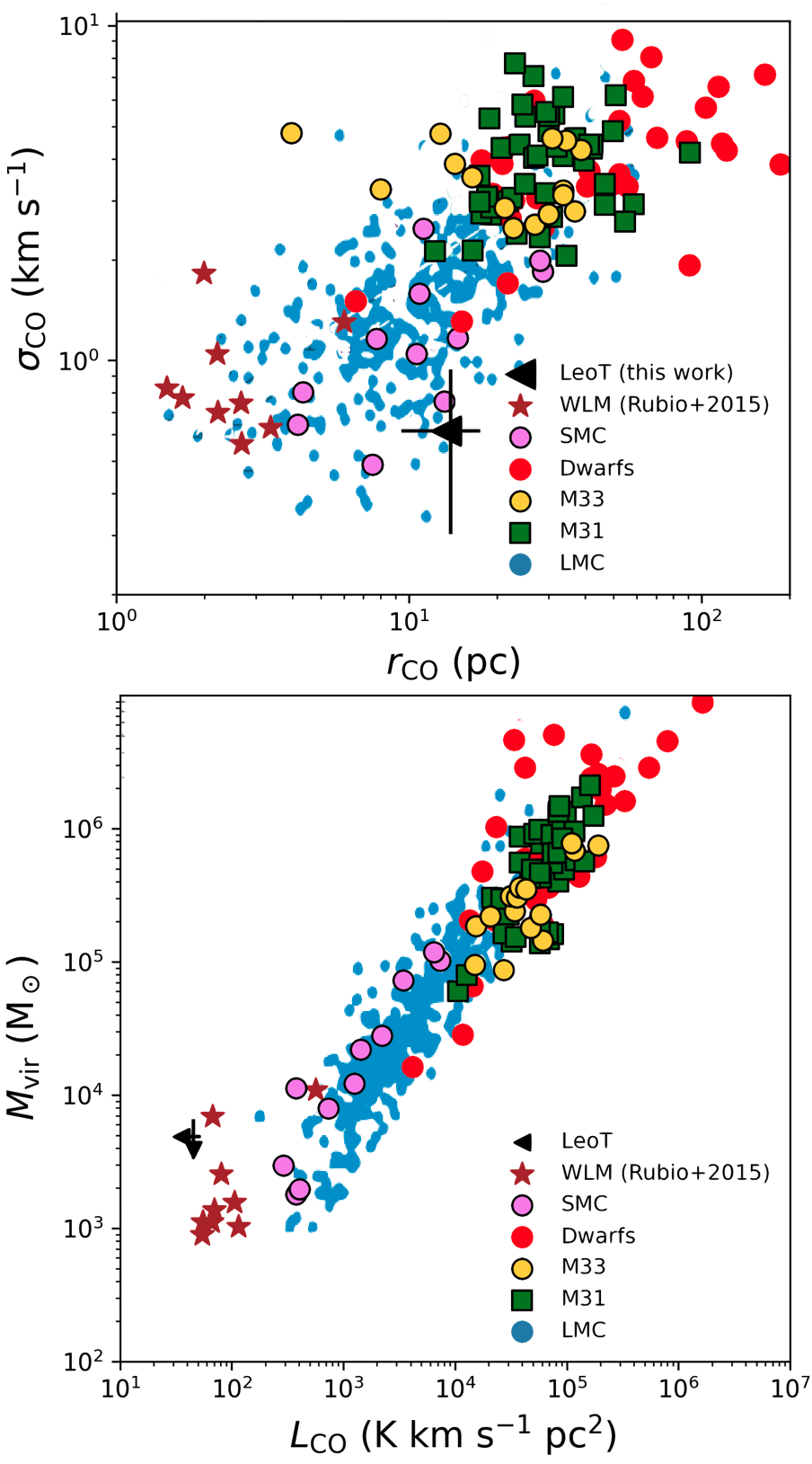}
\vspace{-0.2cm}
\caption{Main properties of the CO clouds identified in this work.
Here are shown the velocity dispersion-size (radius) relation (top panel)
and the virial masses vs. CO(1-0) line integrated luminosities (bottom panel).
The limited angular resolution of the CO data analyzed here allow us to include only an upper limit for $r_{\rm CO}$ (black triangles-arrows) (see Sect.\ref{sec:obs} for more details). The figure also includes results of CO structures detected in other dwarf galaxies: M33 (\citealt{Bolatto2008}); LMC (\citealt{Wong2011}); and a compilation from the literature as included in \cite{Rubio2015} corresponding to WLM, SMC, and M31.}
\label{fig:Lco-Mvir}
\end{figure}

In Fig.\ref{fig:Lco-Mvir} we show the main properties and scaling relations for the clouds in Leo~T and compare them to other galaxies. 
We find that Leo~T clouds have similar luminosities and masses similar to the clouds in the metal-poor Local Group dwarf galaxy WLM \cite{Rubio2015}.
Furthermore, within our resolution, Leo~T clouds appear more extended than in WLM, but consistent within the scatter of the main trends observed in other galaxies that extend between 1 and 100 \pc, as shown in Fig.~\ref{fig:Lco-Mvir}.
Similar results are also reported by \citet{Shi2016} and \citet{Shi2020}, who identify CO structures star-forming regions from galaxies DDO 70 (or Sextans B; $Z_{\rm SextB}\!\sim\! 0.07\!\times\! Z_{\odot}$), DDO 53 ($Z_{\rm DDO53}\!\sim\! 0.14\!\times\! Z_{\odot}$), and DDO 50 ($Z_{\rm DDO50}\!\sim\! 0.18\!\times\! Z_{\odot}$). 
Interestingly, and similarly to WLM (radii$\sim$1.5-6 pc; \citealt{Rubio2015}), they also find that CO clumps in DDO 70 can be up to four times larger (radii$\sim$1.5-2.5 pc) than those in typical massive star-forming regions in the MW ($\sim$0.5-1.0 pc for clumps with $\Sigma_{\rm gas}\!\approx\!700$ M$_\odot$ pc$^{-2}$; \citealt{Shi2020}), suggesting a scenario of suppressed gas fragmentation in the low metallicity regime due to larger Jeans mass, slow cooling, and weak turbulence.

We also place Leo~T in the Kennicutt-Schmidt relation.
For this we estimated the molecular gas densities for the clouds of $\Sigma_{\rm mol}\sim9\sm\pc^{-2}$ (Table \ref{table:obs}). 
Adding \hi and heavier elements result in gas surface mass densities for the North, West and South clouds of $\Sigma_{\rm gas}\!=\!11\sm\pc^{-2}$, $12\sm\pc^{-2}$, and $13\sm\pc^{-2}$, respectively. 
These values agree well with the range found for extremely metal-poor dwarf galaxies \citep[\eg][see their Fig.1]{Filho2016}, where the (log) mean values among the clouds are found 
around $\langle\Sigma_{\rm gas}\rangle\sim\! 10[\sm\pc^{-2}]$ having a scatter ranging between 3 and $100[\sm\pc^{-2}]$. Furthermore, as these are lower limit densities, future high-resolutions observations could potentially find higher densities.
To date, there are only upper limits for the current star formation rate (SFR) in Leo~T, 
where no massive stars, gas-embedded star clusters, HII regions, nor dust have been detected,
where we report no detectable cold dust continuum emission in our ACA data for the targeted sensitivities.
The analysis of H${\alpha}$ emission maps performed by \citet{Vaz2023} determined upper limits for the SFR densities of $\Sigma_{\rm SFR}<10^{-5}\sm\yr^{-1}\kpc^{-2}$,
being much lower than in other metal-poor dwarfs \citep[][]{Filho2016} that show clouds with a (log) mean population value around $\langle\log \left(\Sigma_{\rm SFR}{\sm\yr^{-1}\kpc^{-2}}\right)\rangle\, \approx\! -3$ with a scatter of 1 dex.
Moreover, SFR can vary strongly if Leo~T has bursty or fast episodic cycles of active star formation followed by quiescent episodes, as simulations have predicted for low-mass dwarfs \citep[\eg][]{Read2016}, resulting in non-continuous SFR values.
\citet{Weisz2012} stellar population models of color-magnitude HST data, result in historic star formation rates of $\langle {\rm SFR}\rangle \sim5\times10^{-6}\sm\yr^{-1}$ and even lower in the last 25\Myr.
Adopting the observed total area of the clouds, would result in a star formation density of $\Sigma_{\rm SFR}\sim6\times10^{-3}\sm\yr^{-1}\kpc^{-2}$, which would better agree with the estimates in other dwarfs.
Future ALMA and JWST high-resolution observations could potentially reveal dense cores in the centers of the molecular clouds and provide an estimate of the current SFR.

\section{\hi \, and CO north cloud spectra}
\label{sec:NCloud}
\begin{figure}[ht!]
\hspace{-0.5cm}
\includegraphics[width=9.5cm]{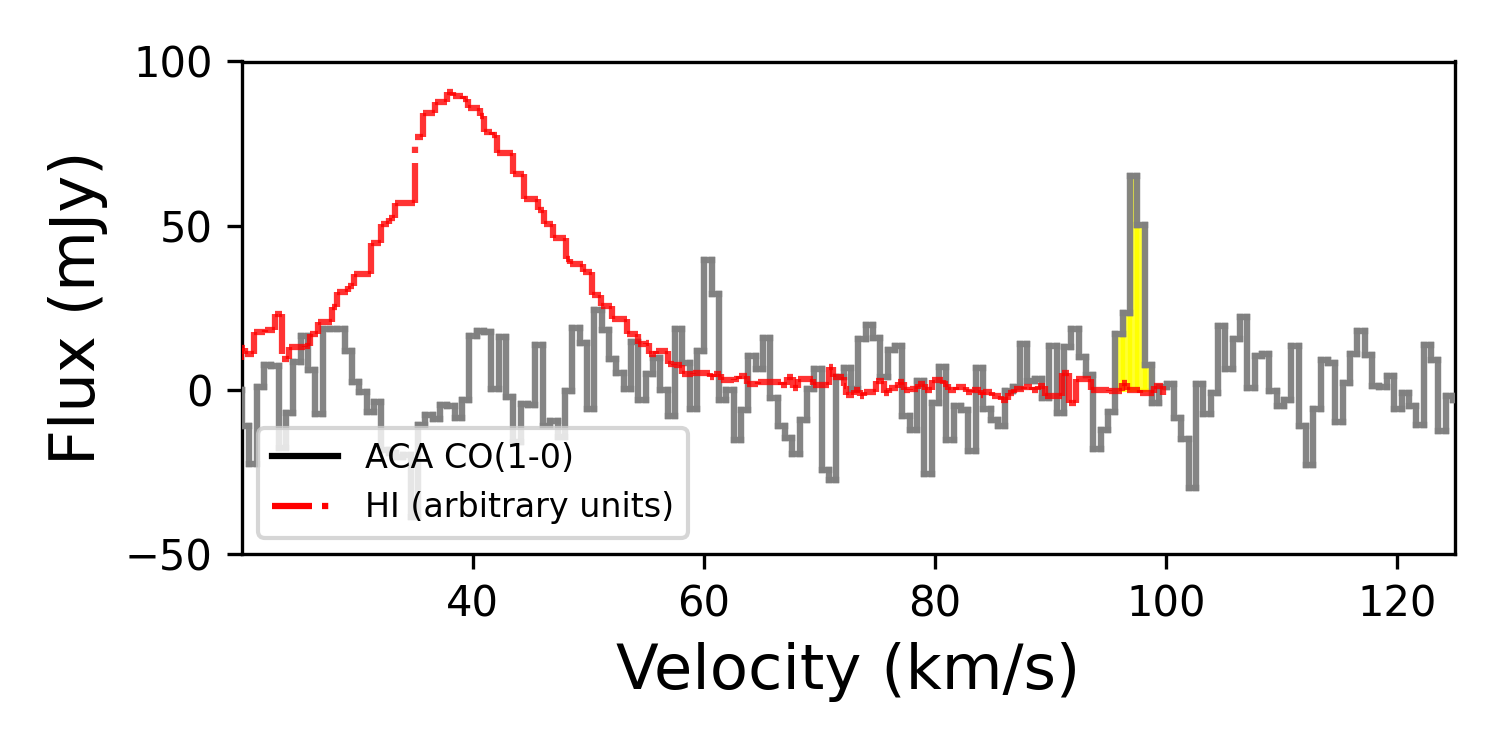}
\vspace{-0.8cm}
\caption{Average \hi spectrum from \cite{Adams2018} (in rescaled units; red dashed line) and this work's CO spectrum of the north cloud (gray solid line), with the detected CO peak (yellow area). 
The velocity offset of the CO emission line peak with respect to the \hi peak ($\sim$60 km s$^{-1}$) suggests the extreme physical conditions that the north cloud is exposed to, which are discussed in Sect.\ref{sec:res}.}
\label{fig:North_spectrum}
\end{figure}

In Fig.\ref{fig:North_spectrum} we show in detail the spectra of the north cloud.

\section{Internal stability of the clouds}
\label{sec:istab}
We also evaluate the internal stability of the clouds, considering resolution limitations, by taking a mean mass density of $\langle\rho_{\rm mol}\rangle\!\sim\! M_{\rm mol}/(\frac{4}{3}\pi r_{\rm CO}^{3})\!\sim\!0.83\sm\pc^{-3}$ to estimate the Jeans masses of the clouds of $M_{\rm J}\!=\!(5\sigma^2 G^{-1})^{3/2}(4\pi\rho/3)^{-1/2}\sim\!5.3\times 10^{3} \sm\gtrsim M_{\rm mol}$; which shows that the clouds would be marginally stable.
In addition, since we apparently detected internal rotation in the molecular clouds (Fig.\ref{fig:co:map}, $v_c\!\!\sim\! |(v_{\rm CO}\! -\! v_{\rm peak})|_{\rm max} \!\sim\!\!0.7$ km s$^{-1}$),
we estimate the Toomre parameter \citep{Toomre1964}, finding an unstable value $Q_{\rm T}\!\approx\!(\sigma v_{\rm c} r^{-1})(\pi G \Sigma)^{-1}\!<\!1$.
We also find free-fall time scales of $t_{\rm ff, mol}\!=\!\sqrt{(3\pi)(32G\rho)^{-1}}\sim9\Myr$.
Future high-resolution observations will better reveal the internal structures in these clouds.

\section{The CO-to-H$_2$ conversion factor in Leo~T and in foreground MW material}
\label{sec:co2h2}

\begin{figure}[ht!]
\hspace{-0.5cm}
\includegraphics[width=9.6 cm]{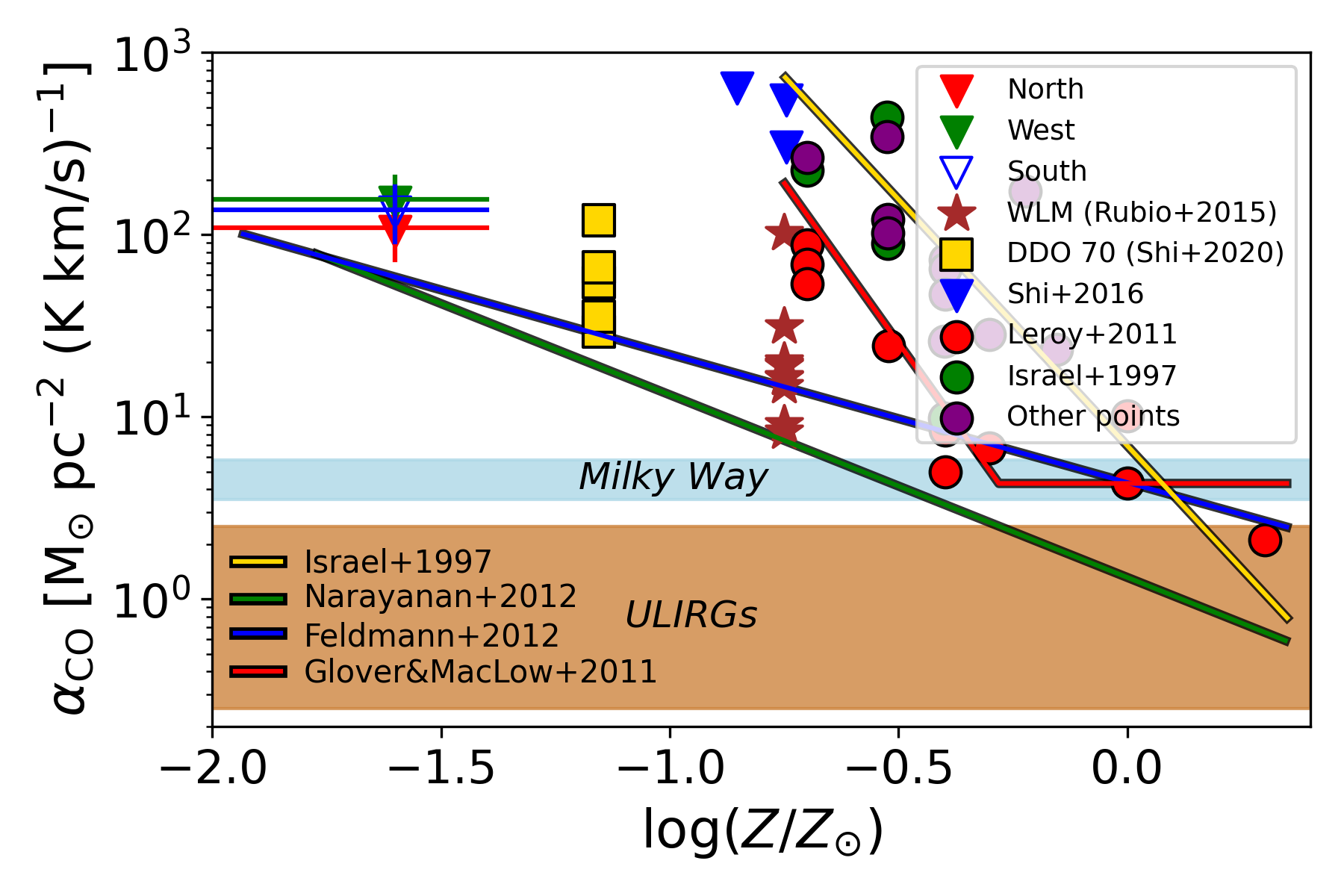}
\vspace{-0.7cm}
\caption{
CO-to-H$_2$ conversion factor ($\alpha_{\rm CO}$) vs. metallicity, adapted from \cite{Bolatto2013a} to include the main properties of the CO clouds in Leo~T detected in this work. The $\alpha_{\rm CO}$ estimations (red, green and unfilled-red inverted triangles) correspond to upper limits (see Sect.\ref{sec:res}), with uncertainties of the order of $\sim$24-33\%. The figure includes a recompilation of CO structures reported in different galaxy samples: WLM \citep{Rubio2015}; DDO 70 \citep{Shi2020}; DDO 50 and DDO 53 \citep{Shi2016}; local LIRGs and ULIRGs \citep{Leroy2011}; LMC and other magellanic irregular galaxies \citep{Israel1997}; and other local group galaxies \citep{Madden1997,Gratier2010,Smith2012}. The figure also contains theoretical models of $\alpha_{\rm CO}(Z)$ from \cite{Israel1997} (yellow line), \cite{Narayanan2012} (green line), \cite{Feldmann2012} (blue line), and \cite{Glover&MacLow2011} (red line). The light blue and brown areas correspond to the $\alpha_{\rm CO}$ ranges for the MW and ULIRGs recommended by \cite{Bolatto2013a}.
}
\label{fig:alpha}
\end{figure}

We determined the CO($J$=1-0)-to-H$_2$ conversion factor $\alpha_{\rm CO}$ for the three molecular clouds by measuring the dynamical masses and CO luminosities within the circular areas of radius $r_{\rm CO}$, resulting in upper limits due to resolution limitations. Future follow-up observations could potentially reveal a more compact mass distribution or even a subclump substructure with smaller compact cores (see Fig. \ref{fig:Lco-Mvir}).
In Figure \ref{fig:alpha} we compare the $\alpha_{\rm CO}$ values of the clouds with similar studies, such as \citet{Rubio2015}, where they find high values for the WLM galaxy with an average of $\alpha_{\rm CO}$$\sim$28 M$_\odot$ (${\rm K\, km\, s^{-1}\, pc^2}$)$^{-1}$ and the highest having $\alpha_{\rm CO}$$\sim$124 M$_\odot$ (${\rm K\, km\, s^{-1}\, pc^2}$)$^{-1}$.
They suggested that this extreme $\alpha_{\rm CO}$ value may reflect the almost null CO emission for most of the cloud volume filled by H$_2$; this is because CO resides primarily in the densest cores of the H$_2$ clouds. 
Similar results are also reported by \cite{Shi2016} and \cite{Shi2020}, who identify CO structures star-forming regions from galaxies DDO 70 (or Sextans B; $Z_{\rm SextB}\!\sim\! 0.07\!\times\! Z_{\odot}$), DDO 53 ($Z_{\rm DDO53}\!\sim\! 0.14\!\times\! Z_{\odot}$), and DDO 50 ($Z_{\rm DDO50}\!\sim\! 0.18\!\times\! Z_{\odot}$).
They cover $\alpha_{\rm CO}$ (core and cloud) values in the range between 30 and 120 M$_\odot$ (${\rm K\, km\, s^{-1}\, pc^2}$)$^{-1}$ for CO clouds identified in the dwarf galaxy DDO 70 (yellow squares in Fig.\ref{fig:alpha}; \citealt{Shi2020}). 
They argue that since $\alpha_{\rm CO}$ increases with decreasing metallicity, in combination with the low metallicities ($\sim$0.07-0.18$Z_\odot$) and the large sizes (1.5-3 pc) of the CO clouds analysed, may result in suppressed gas fragmentation.
The latter is a consequence of large clumps not being able to further fragment into small ones due to large Jeans mass and weak turbulence.
Moreover, we note that the $\alpha_{\rm CO}$ estimates in Fig.\ref{fig:alpha} are calculated using different methodologies and at different scales, with \citet{Rubio2015,Shi2020} showing $\alpha_{\rm CO}$ values of compact CO cores using a virial estimator similar to our approach used for the Leo'T clouds that could appear more extended clouds due to resolution limitations, while \citet{Rubio2015} also include $\alpha_{\rm CO}$ values estimated with dust to gas ratios.
More recently, \cite{Komugi2023} report a weak CO detection in the star-forming regions in DDO 154 (12+log[O/H]=7.67), estimating a lower limit of $\alpha_{\rm CO}\gtrsim 10^3$ M$_\odot$ (${\rm K\, km\, s^{-1}\, pc^2}$)$^{-1}$ derived from  H$\alpha$-based star formation rates \citep{Verter&Hodge1995,Taylor&Klein2001}.
Moreover, using dust continuum data to measure the molecular gas content in DDO 154, they also estimate a cloud $\alpha_{\rm CO}$ factor of at least two orders of magnitude higher than that of the MW ($\sim$4.3 M$_\odot$ [${\rm K\, km\, s^{-1}\, pc^2}$]$^{-1}$; e.g., \citealt{Walter2008}).

Furthermore, in Sect.\ref{sec:res}, we mentioned that the $\Delta \upsilon_{\rm CO-HI}\!=\!+57.7\!\pm0.7$  km s$^{-1}$ offset of the north CO cloud (see Fig.\ref{fig:North_spectrum}) could also imply that this cloud could belong to MW foreground material where \hi transitions to H$_2$. From the HI4PI survey data of Galactic \hi emissions \citep{Westmeier2018}, we find heliocentric velocities towards Leo~T's direction of $\upsilon^{\odot}_{\rm los,\hi}\sim80-90$ km s$^{-1}$, which are consistent with the North cloud ($\upsilon^{\odot}_{\rm los,\,CO}\sim$97 km s$^{-1}$).
Assuming that the north cloud is within the \hi MW disk diameter ($D_{\rm HI,MW}$$\sim$40 kpc; \citealt{Westmeier2018}), we can compute a new CO luminosity and mass of $L'_{\rm CO}$=$\sim$0.41 K km s$^{-1}$ pc$^{2}$ and $M'_{\rm mol}$$\sim$460 M$_{\odot}$ for the north cloud, respectively, yielding an $\alpha'_{\rm CO}$$\sim$1.12 $\times 10^3$ M$_\odot$ pc$^{-2}$ [K km s$^{-1}$]$^{-1}$. 
However, this value is extremely high compared to the typical $\alpha_{\rm CO}$ values in the MW (Fig.\ref{fig:alpha}). 
For example, using the measurements of $\alpha_{\rm CO}$ and CO luminosities in the MW by \cite{Solomon1987}, \cite{Bolatto2013a} derived the empirical relation $\alpha_{\rm CO}\!=\!(L_{\rm CO}/10^5)^{-0.185}$. For the new $L'_{\rm CO}$, this results in a $\alpha_{\rm CO}$$\sim$10 M$_\odot$ pc$^{-2}$ [K km s$^{-1}$]$^{-1}$, which is two orders of magnitude lower than $\alpha'_{\rm CO}$. In consequence, the physical quantities of the north cloud are more plausible when we assume that it is located at the distance of Leo~T.

\section{List of Abbreviations}
\label{sec:list}
AGB: Asymptotics Giant Brach, 
AGNs: Active Galactic Nuclei,
ALMA: Atacama Large Millimeter Array,
CASA: Common Astronomy Software Applications,
CDM: Cold-Dark-Matter, 
CGM: circumgalactic medium,
IGM: intergalactic medium,
ISM: interstellar medium,
HST: {\it Hubble} Space Telescope,
WLM: Wolf–Lundmark–Melotte, 
WSRT: Westerbork Synthesis Radio Telescope.

\section{Acknowledgements}

The authors deeply thank Elizabeth Adams and Tom Oosterloo for making the HI data available.
M.B. thanks C.Agurto for helpful discussions.
V.V. acknowledges support from the ANID BASAL project FB210003 and from ANID - MILENIO - NCN2024\_112. 
M.R. acknowledges partial support from ANID (CHILE) through Basal FB210003. 
D.C. acknowledges funding from the Alexander von Humboldt Foundation.
This paper uses the ALMA data: ADS/JAO.ALMA \#2024.1.00951.S. 
The Westerbork Synthesis Radio Telescope is operated by ASTRON, the Netherlands Institute for Radio Astronomy, with support from the Netherlands Foundation for Scientific Research (NWO).
The authors deeply thank the citizens of Chile for their tax contributions that allow the national development of science and this project.

\end{document}